\documentclass[preprint2]{aastex}
\usepackage{graphicx}

\newcommand{\be}{\begin{equation}}
\newcommand{\ee}{\end{equation}}
\newcommand{\bea}{\begin{eqnarray}}
\newcommand{\eea}{\end{eqnarray}}
\newcommand{\gtabouteq}{\,\hbox{\raise 0.5 ex \hbox{$>$}\kern-.77em 
                    \lower 0.5 ex \hbox{$\sim$}$\,$}}       
\newcommand{\ltabouteq}{\,\hbox{\raise 0.5 ex \hbox{$<$}\kern-.77em 
                     \lower 0.5 ex \hbox{$\sim$}$\,$}}

\shorttitle{Uncovering AGNs in Polarization}
\shortauthors{Irwin et al.}

\begin{document}


\title{CHANG-ES VIII: \\
Uncovering Hidden AGN Activity in Radio Polarization
}


\author{
  Judith A. Irwin\altaffilmark{1},
  Philip Schmidt\altaffilmark{2},
  A. Damas-Segovia\altaffilmark{2},
Rainer Beck\altaffilmark{2},
Jayanne English\altaffilmark{3}
George Heald\altaffilmark{4}
Richard N. Henriksen\altaffilmark{1},
Marita Krause\altaffilmark{2},
Jiang-Tao Li\altaffilmark{5},
Richard J. Rand\altaffilmark{6},
Q. Daniel Wang\altaffilmark{7},
Theresa Wiegert\altaffilmark{1},
Patrick Kamieneski\altaffilmark{7},
Dylan Par{\'e}\altaffilmark{7}, and
Kendall Sullivan\altaffilmark{7}
}
\altaffiltext{1}{Dept. of Physics, Engineering Physics, \& Astronomy, Queen's University,
    Kingston, Ontario, Canada, K7L 3N6 {\tt irwin@astro.queensu.ca, henriksn@astro.queensu.ca, twiegert@astro.queensu.ca }.}
\altaffiltext{2}{Max-Planck-Institut f{\"u}r Radioastronomie,  Auf dem H{\"u}gel 69,
53121, Bonn, Germany,
{\tt pschmidt@mpifr-bonn.mpg.de,
  adamas@mpifr-bonn.mpg.de, rbeck@mpifr-bonn.mpg.de, mkrause@mpifr-bonn.mpg.de}.} 
\altaffiltext{3}{Department of Physics and Astronomy, 
  University of Manitoba, Winnipeg, Manitoba, Canada, R3T 2N2 {\tt jayanne\_english@umanitoba.ca}.}
\altaffiltext{4}{CSIRO Astronomy and Space Science, 26 Dick Perry Avenue, Kensington WA 6151, Australia;
Netherlands Institute for Radio Astronomy (ASTRON), 
Postbus 2, 7990 AA, Dwingeloo, The Netherlands {\tt George.heald@csiro.au}}
\altaffiltext{5}{Department of Astronomy, University of Michigan, 409 West Hall, 1085 S. University,
Ann Arbor, MI, 48109 {\tt jiangtal@umich.edu}.}
\altaffiltext{6}{Dept. of Physics and Astronomy, University of New Mexico, 
1919 Lomas Boulevard, NE, Albuquerque, NM, 87131, USA {\tt rjr@phys.unm.edu}.}
\altaffiltext{7}{Dept. of Astronomy, University of Massachusetts, 710 North
Pleasant St., Amherst, MA, 01003, USA, 
{\tt wqd@astro.umass.edu, pkamieneski@umass.edu,
dpare@umass.edu,
kendallsulli@umass.edu.}}




\begin{abstract}
  We report on C-band (5 - 7 GHz) observations of the galaxy, NGC~2992, from
  the CHANG-ES sample.  This galaxy displays an embedded nuclear double-lobed radio morphology
  within its spiral disk, as revealed in linearly polarized emission but {\it not} in total intensity
emission. 
The radio lobes are kpc-sized,
similar to what has been observed in the past for other Seyfert galaxies, and show ordered
magnetic fields. NGC~2992 has shown previous evidence
for AGN-related activity, but not the linearly polarized
radio features that we present here.
We draw attention to this galaxy as the first clear example (and prototype) of bipolar radio outflow that is
revealed in
linearly polarized emission only.
Such polarization observations, which are unobscured by dust,
provide a new tool for uncovering hidden weak AGN activity which
 may
 otherwise be masked by brighter unpolarized emission within which it is embedded.
The radio lobes
observed in NGC~2992 are interacting with the surrounding interstellar medium and
offer new opportunities to investigate the interactions between nuclear outflows and the ISM in
nearby galaxies. 
 We also compare the radio emission with a new
 CHANDRA X-ray image of this galaxy. A new CHANG-ES image of
 NGC~3079 is also briefly shown as another example as to how much more obvious
 radio lobes appear in linear polarization as opposed to total intensity.

\end{abstract}





\section{Introduction}
\label{sec:introduction}

The powerful jet and double-lobed character of radio emission around distant massive
elliptical galaxies is well known and 
originates in the active galactic nuclei (AGNs)
of their host galaxies. Indeed, since
it is believed that the formation of a supermassive black hole (SMBH) is closely
connected to the formation of the bulge mass, with both evolving together
\citep{fer00,geb00,zhe09,kor13}, one would expect AGN-related activity
to manifest itself in spirals as well. Yet,
spiral
galaxies do not generally show such impressive evidence for nuclear activity
and observations of AGNs in spirals are relatively rare.  
The incidence of radio AGN in late type galaxies, for instance, is about an order of magnitude 
lower than in early-type galaxies \citep{kav15b}.  Those spiral galaxies that
do display nuclear outflows in the form of 
 AGN-related jet and/or
lobe-like features make for a short list.  For example,
Seyfert galaxies can harbour jets
on pc to kpc scales \citep[][and references therein]{ho01}.  Well-known examples
also
include NGC~3079 \citep{irw88,hum83}, the Circinus spiral galaxy \citep{har90,gre03},
NGC~7479 \citep{lai08}, NGC~4388 \citep{dam16},
NGC~4258 \citep{gre95,miy95,her98,kra04}, as well as the spectacular examples,
0313-192 \citep{led01,kee06}, `Speca' \citep{hot11},
J2345−0449 \citep{bag14},
and J1649+2635 \citep{mao15}.

The reason for the apparent relative scarcity of such activity in spirals
is not entirely clear, though a variety of
possibilities have been raised.  For example, advection-dominated accretion flow
may result in very little radiation \citep{nar95,yua14}, the SMBHs in spirals may not be massive
enough to produce the large-scale jets seen in ellipticals \citep{lao00},
accreting SMBHs in spirals may not be spinning rapidly enough \citep{wil95,sik07,tch10,dot13}
or the magnetic flux could be insufficient to launch jets \citep{sik13}. 
 Other issues such as
the effect of the dense surrounding 
 interstellar medium (ISM) on the outflow \citep{das15}, 
the difficulty in accreting high angular momentum gas, the possibility
of episodic outflows \citep{mar03}, and
environmental effects such as perturbations from nearby galaxies,
 all have yet to be fully understood and quantified.

An additional complicating issue is that nuclear
outflows can also originate from a central starburst
\citep[][and references therein]{vei05}
and both may be occurring in the same galaxy
\citep[e.g. NGC~3079][and references therein]{sha15}.  
It has thus been difficult to identify the origin of an observed outflow as being due to
star formation (SF), or an AGN, or some combination thereof. 
An active low-luminosity AGN (LLAGN), for example, could be embedded in a nuclear region with
abundant SF activity.
Since all spirals host
star formation, the issue is really whether or not an AGN is also present and 
how best to identify it.

Hard X-ray emission is an indicator of AGN activity,
provided the luminosity exceeds what is expected from ultra-luminous X-ray
sources ($10^{39-42}$ ergs/s) \citep{mus04}. 
Optical emission line ratios that indicate the
presence of a hard ionizing continuum is another \citep{bal81}
although dust obscuration can be problematic, especially for edge-on galaxies.
The mid-IR PAH\footnote{Policyclic Aromatic Hydrocarbon} to continuum ratio 
has been used as a SF/AGN diagnostic for distant galaxies \citep{kir15}
and other mid-IR line ratios \citep[see][]{per10} have also been shown to be effective.

Evidence for an AGN, however, is not necessarily evidence for AGN-related outflows.
As indicated above, it is in the radio continuum that jets
and lobes have historically been detected, since
synchrotron emission is a strong component of such outflow. 
AGN-related outflows have been identified by
the presence of linear structures in Seyfert galaxies \citep{ulv84} or by small-scale jets in late-type spirals
\citep{kav15a}, but only in 
in {\it total intensity} radio
continuum \citep{ulv84}.
In order to be so identified, such emission must
dominate any SF-related emission in the vicinity.

In this paper, we provide a new diagnostic for identifying AGN-related outflow via
radio polarization, specifically {\it bipolar radio features or other evidence for AGN-related
  outflows that
  are masked in total intensity but are revealed in linear polarization.}  In other words, LLAGNs may
not, in fact, be rare, but require that our observations be targeted at the polarized rather than the total
intensity radio emission
to be clearly seen.

 Polarization images are well-suited to uncovering previously hidden AGNs
 because of their sensitivity to ordered magnetic fields that are prevalent in the outflows
 and also because the radio emission is insensitive to dust obscuration.
As part of the CHANG-ES program  \citep[`Continuum Halos in Nearby Galaxies -- an EVLA
  Survey',][]{irw12a}, we focus on one such galaxy at C-band (6 GHz), i.e. NGC~2992, which
we take to be a prototype for such activity.  This galaxy
was previously known to harbour an active nucleus (Sect.~\ref{sec:n2992}) but now the
polarized radio emission clearly shows the outflows.


Note that NGC~2992 is {\it not} the only AGN in the CHANG-ES sample.
A more complete census of
AGNs in the CHANG-ES sample will be presented in a future paper. 
For a list of radio polarization observations and magnetic fields that have
been observed in other mostly non-edge-on
nearby galaxies, see the appendices of \citet{bec13}.

\section{NGC~2992 and its AGN}
\label{sec:n2992}

\begin{figure*}[h]
   \includegraphics*[width=1.0\textwidth]{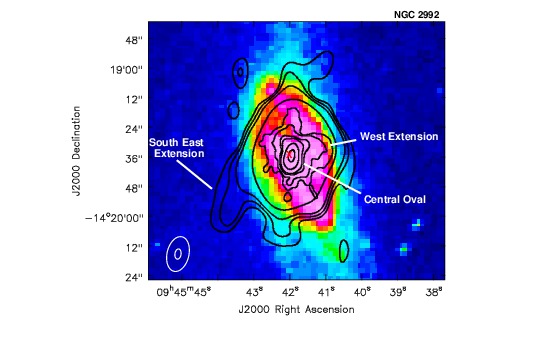}
   \hspace{-1.20in}
   \caption{C-band total intensity contours on a Digitized Sky Survey (DSS2) blue image of NGC~2992. The four
outermost contours are D-configuration data with contour levels at 0.05 (6$\sigma$), 0.07, 0.1, and 0.3 mJy beam$^{-1}$;
the peak D  specific intensity is 65.6 mJy  beam$^{-1}$.  The inner six contours are C-configuration data with
levels of 0.035 (10$\sigma$), 0.1, 0.5, 1.0, 5.0, and 20 mJy beam$^{-1}$; the peak C-configuration specific intensity is
28.0 mJy beam$^{-1}$.  The two beam sizes, both robust 0 weighting (Table~\ref{table:image_parameters}) are shown at lower left
and the galaxy NED center is marked with a red X. Several features mentioned in the text are labelled.
}
\label{fig:n2992_totalintensity_DSS}
\end{figure*}

NGC~2992 
\citep[Fig.~\ref{fig:n2992_totalintensity_DSS}, D = 34 Mpc,][]{wie15} is
a well-known Seyfert galaxy
\citep{osm74, war80} that is interacting with its primary companion, NGC~2993, $\sim 3$ arcmin to the SE 
as well as Arp~245N $\sim 2$ arcmin to the NE (see also Fig.~\ref{fig:combined_optical}).
\citet{duc00} have shown that an additional smaller galaxy, FGC~0938, 
 $\sim 5$ arcmin to the SW is also part of this system, collectively known as Arp~245.
The result is that NGC~2992 is highly disturbed.  For example, Duc et al. list the galaxy's inclination at
70 degrees.  However, in the central $\sim$ 1 arcmin where we detect
continuum emission, optical isophotes\footnote{Measured from the DSS2 blue
  image since an SDSS image is not available.}
suggest an inclination of 64 degrees assuming a thin disk.
Note that the near side of the galaxy
 is the west side \citep{vei01}.

A variety of evidence has already pointed to nuclear activity in this galaxy.  A large scale radio feature 
extends $\approx$
30 arcsec (5 kpc) to the SE \citep{hum83, war80} which we also detect 
in total intensity (Fig.~\ref{fig:n2992_totalintensity_DSS}) and refer to as the `southeast extension'.
Higher resolution radio continuum observations by \citet{ulv84} and \citet{weh88} show emission
within the central 10 arcsec of the galaxy in the form of a double loop that is 
limb-brightened and centered on the nucleus.  Our data do not resolve the double-loop and instead this region
appears as a
central oval-shaped region  at
a position angle of 170 degrees in Fig.~\ref{fig:n2992_totalintensity_DSS} (the `central oval').
  \citet{cha00} suggest that
  multiple radio components are superimposed in this region.
  Hard mid-IR line ratios of 2.1 and 2.8 for
  [NeII](12.81 $\mu$m)/[NeV](14.32 $\mu$m) and [OIV](25.89 $\mu$m)/[NeII](12.81 $\mu$m),
  respectively, place NGC~2992 firmly in the Seyfert/AGN part of the diagnostic plots of
  \citet{per10} (their Figs. 5 and 6).

Evidence for nuclear radial outflows have also been found in optical emission lines  
\citep{col87,tsv95, col96, mar98,vei01}; detailed analyses of optical emission lines 
can be found in \citet{all99} and \citet{gar01}. 
Its Seyfert classification has changed with time between 1.5 and 2 within a
  few years \citep{tri08}.
 The optical outflows have been attributed predominantly
 to the AGN rather than a nuclear starburst \citep{fri10}. 

NGC~2992 has long been known to be an X-ray source and the X-ray emission varies 
\citep{mac82,elv90,kru90,wea96,gil00,beckm07,shu10} (flares) on timescales
of days to weeks \citep{shu10}. 
The companion galaxy, NGC~2993, is also an X-ray source \citep{col98}.
An infrared outburst was detected in 1988 \citep{gla97}.  \citet{col05} conclude
that the extra-nuclear soft X-ray luminosity could be explained either
by a star-burst-driven or AGN-driven wind.

In summary, there is compelling prior evidence for an AGN in NGC~2992.
In total intensity radio emission, however, only the eastern extension and
inner double-loop (within our central oval) have previously
indicated that
AGN-related outflow has occurred.


\section{Observations and Data Reductions}
\label{sec:obs_data}

\subsection{Individual Configurations}
\label{sec:individual_configurations}

Our CHANG-ES data include both C-band (6 GHz) and L-band (1.6 GHz) observations
using the  Karl G. Jansky Very Large Array (hereafter the VLA) in a variety of array
configurations.  Here, we report on the 
C-band observations
only, since no linear polarization is observed at L-band with a fractional polarization (P/I)
that is above $0.5\%$ in either
array configuration; this is likely due to Faraday depolarization at the lower frequency.
The C-band observations were obtained in the C and D configurations
of the VLA.
The largest spatial scale detectable by the VLA at C-band is 4 arcmin; since the entire field
of view shown in 
Fig.~\ref{fig:n2992_totalintensity_DSS} is only 1.6 arcmin, we should not have missed
any significant short-spacing flux.
A summary of the
observations and image parameters is given in Table~\ref{table:image_parameters}.

Data reduction and imaging details are described in \citet{irw13} and \citet{wie15}.  Here we provide a
brief
outline.  The Common Astronomy Software Applications (CASA)
 package \citep{mcm07} was used throughout\footnote{\tt http://casa.nrao.edu}.

Each galaxy was observed within
a scheduling block that included other galaxies so that the uv coverage could be maximized for each. 
The calibrators were 3C~286 
which was used for the primary gain and phase calibration as well as the bandpass and polarization
angle calibration, J1407+2827 (OQ208) which was used to calibrate
the polarization leakage, and finally the secondary gain and phase calibrators, J0837-1951 and
J0943-0819 for NGC~2992 in D and C configurations, respectively.


The frequency coverage was from 4.979 $\rightarrow$ 7.021 (2.042 GHz) and observations were carried
out in spectral line mode with 16 spectral windows across the band (each 128 MHz wide).  Each spectral window
contained 64 spectral channels for a total of 1024 channels across the entire band.
  
Flagging was carried out manually 
and any flagging that applied to the total intensity
data (RR and LL circular polarizations) was automatically applied to the cross-polarizations
(RL and LR). Additional flagging was then carried out for RL and LR as required.

Wide-field imaging \citep{cor08b} was carried out using the
multi-scale multi-frequency synthesis (ms-mfs) algorithm
\citep{cor08,rau11}, with simultaneous fitting of a simple power law 
($S_\nu\,\propto\,\nu^\alpha$) across the band. 
For all maps from the individual observing runs, Briggs robust = 0 uv weighting was used \citep{bri95}
as implemented in CASA\footnote{A second set of maps using a uv taper was also made for internal comparison but
 are not presented here.}.
Self-calibration iterations were also carried out for the total intensity image (Table~\ref{table:image_parameters})
until no further improvement 
in the dynamic range resulted.

Imperfect phase and amplitude calibrations produce errors that 
normally be below the rms for 

The dynamic range of the total intensity images are very high 
(e.g. $\approx$ 8000/1 at C-configuration) and therefore imperfect phase and amplitude calibrations
result in some residual sidelobes
 that could not be cleaned out completely.
Hence we show our first contours at levels that are higher than is normally shown
(Fig.~\ref{fig:n2992_totalintensity_DSS}).


 For each observation,  Stokes Q and U were imaged
using the same self-calibration table and the same
 input parameters as for total intensity. 
 Maps of linear polarization, $P\,=\,\sqrt{Q^2\,+\,U^2\,-{\sigma_{Q,U}}^2}$ were then made, where $\sigma_{Q,U}$ is the rms
 noise of the Q and U maps.  The latter term does a zeroth order correction for 
 the fact that P images are positively biased 
 \citep[e.g.][]{sim85,vai06,eve01} and is the only correction that is currently implemented in CASA. In this
correction some very
low level emission becomes negative and those negatives are then blanked.
Polarization angle maps were then
also made with a 3$\sigma_{Q,U}$ cut off.

All maps were then corrected for the primary beam (PB, full-width at half-maximum = 7.5 arcmin).
We show our resulting images without the PB
correction (unless otherwise indicated -- see next section)
so that the noise is uniform; however, any measurements are made on the
PB-corrected images, and so throughout. Note that the source is only about 1 arcmin in size so that PB corrections are
minor\footnote{Recent PB measurements are now available for the VLA (EVLA Memo 195, R. Perley)
  though (at the time of writing) have not yet been implemented into CASA; the FWHM is about 6\% smaller than previously thought
  at C-band.}.




\subsection{Combined Configurations and Rotation Measure (RM) Synthesis}
\label{sec:combined_configurations}

We combined the C and D configuration uv data
into a single large data set in order to
obtain the highest possible signal-to-noise (S/N) ratio with sensitivity to a wide range of spatial scales.
 In addition, we carried out an RM Synthesis analysis to correct for
Faraday rotation, as follows.

Both the C and D configuration uv data were imaged simultaneously to make two new images of
`low resolution' and `high resolution' which resulted from modifying the Briggs robustness parameter and
uvtaper
(Table~\ref{table:image_parameters}).
The total intensity images were made in the same way as described in Sect.~\ref{sec:individual_configurations}
except for the different input parameters which are listed
in the table\footnote{The {\it multiscales}
  parameters were also slightly different from the individual configurations, but 
   they matched between the total intensity and corresponding polarized intensity images.}.
The original images had marginally smaller beams than quoted in the table and were then smoothed
to the resolutions indicated; this was to ensure that the total intensity and polarized
intensity resolutions matched exactly since they are formed using different processes (see below).
Attempts were made to do additional self-calibration of the combined configuration data sets; however
such attempts did not improve the dynamic range of the images.  

For the linear polarization imaging, we produced combined-configuration Q and U cubes with dimensions of
RA, DEC, and frequency.  For each image plane along the frequency axis, we averaged
the data over each spectral window so that we sample the entire band with a new channel
separation of 128 MHz.  Images for each new channel were then smoothed  to the resolution
of the spectral window with the lowest resolution and then
PB-corrected using the primary
beam that corresponds to the center of its channel.

Stokes Q, U and polarized intensity (PI) cubes in dimensions, RA, DEC, Faraday depth, $\phi$,
were then generated using the RM synthesis code of \citet{bre05}. The Faraday depth of a source is defined as
\begin{equation}
\phi(\vec r)\,=\,0.81{\int_{\vec r}^{0}}n_e {\vec B} \cdot d{{\vec {r^{\prime}}}}~~{\rm rad ~m^{-2}}
\end{equation}
where $B$ is the magnetic field strength in $\mu$G, $n_e$ is the electron number density in cm$^{-3}$, and
r is the line-of-sight path length in pc. The $\phi$ axis
of the cubes extends from -1000 to +1000 rad m$^{-2}$ so that $\phi \,\delta \lambda^2\,<<\,1$ is 
fulfilled (where we assumed $\delta \lambda^2$
to be the difference in wavelength squared that corresponds to the separation between
the two lowest frequency channels) and hence no significant Faraday depolarization between
adjacent frequency channels occurs.

The following data products were then extracted from the Faraday cubes via an algorithm kindly supplied by
B. Adebahr (private communication): a) a PI map, consisting of the maximum value along the $\phi$ axis of the PI cube
for each pixel, b) a map of the $\phi$ value at which the maximum PI occurs, c) a map of apparent
polarization position angles corresponding to the $\phi$ value at which the maximum PI occurs, and
d) a map of de-rotated (corrected) polarization position angle which also corresponds to the location at which PI is maximum.
Maps that show $\phi$ at the highest polarized flux density in the Faraday depth spectra, and of the intrinsic
polarization angle for this $\phi$, were computed only in regions where the polarized intensity is
higher than 4$\sigma$ in the Faraday Q and U image planes.

The width of the $\lambda^2$ distribution probed in C-band leads to a RM spread function with a main peak
FWHM of ~2000 rad m$^{-2}$, which implies that potentially present emission at multiple Faraday depths is not
resolved. We therefore assume for our further analysis that emission on any given line of sight
appears at a single Faraday depth. While it would hence be possible to measure Faraday depths at a similar
accuracy by computing 'classical' RMs between different sections of the frequency band (the $n\pi$ ambiguity is
not a major issue at C-band), our main purpose of applying RM synthesis instead is to improve our sensitivity
to the polarized emission.  Moreover, due to the poor resolution in Faraday depth, a deconvolution of
the Faraday spectra was not necessary.

Since maps made as a result of RM synthesis must necessarily be corrected for the PB at a prior stage, any
combined-configuration polarization images are shown with PB correction. Rms noise values are then a function of
position on the map and we quote averages in Table~\ref{table:image_parameters} in these cases.


{
\begin{deluxetable}{lccccccccc}
\hspace*{-10cm}
\tabletypesize{\scriptsize}
\renewcommand{\arraystretch}{0.9}
\tablecaption{C-band Observations and Imaging of NGC~2992\label{table:image_parameters}}
\tablewidth{0pt}
\tablehead{
  \colhead{Observation}  & \colhead{Date\tablenotemark{a}} & \colhead{$\nu_0$\tablenotemark{b}}
  & \colhead{uv Weighting\tablenotemark{c}}
& \colhead{Beam parameters\tablenotemark{d}} &
 \colhead{Pixel size} & \colhead{SC Iterations\tablenotemark{e}}
&\multicolumn{2}{c}{rms noise\tablenotemark{f}} \\
 &        &              &  &             &  & & I & Q,U \\
 & & (GHz) & & $\prime\prime$, $\prime\prime$, deg. &  $\prime\prime$
 & &\multicolumn{2}{c}{($\mu$Jy beam$^{-1}$)}\\ 
}
\startdata
{\bf Individual}\\
\tableline
D configuration & 13-Dec-2011  & 6.000 & Rob=0 & 14.33, 8.84, -13.3 &  1.0 & 2A\&P & 8.4 & 8.0 \\
C configuration & 07-Apr-2012 & 6.000 & Rob=0 & 4.12, 2.48, -8.0&  0.5 & 4P+1A\&P & 3.5 & 3.2 \\
\tableline
{\bf Combined}\\
\tableline
\multicolumn{2}{l}{C+D configuration (low resolution)}    & 6.000 & Rob=+2 (22 klambda) & 10.0,10.0,0.0 &  0.5 & &  15.0  & 3.5 & \\
\multicolumn{2}{l}{C+D configuration (high resolution)}     & 6.000 & Rob=-2 & 5.20, 3.10, -9.5 & 0.5 &  & 5.0 & 7.7 &  \\
\enddata
\tablenotetext{a}{Date that the observations were carried out (UTC).}
\tablenotetext{b}{Central frequency of each image. 
  The bandwidth, before flagging, was 2 GHz. }
\tablenotetext{c}{Briggs `Robust' weighting as implemented in CASA. Value in parentheses indicates
the value of the uv taper for the low resolution combined configuration image.}
\tablenotetext{d}{Major \& minor axis diameters and position angle of the synthesized beam.}
\tablenotetext{e}{Number of self-calibration iterations that were applied to improve the maps.
  Each iteration acted on the non-self-calibrated data, but using
an improved model.  P: phase-only, A\&P: amplitude and phase. }
\tablenotetext{f}{Rms noise for the total intensity (I) and
  Stokes Q and U maps. For Q and U of the combined configuration maps for which RM synthesis was carried out,
  these are averages of the PB-corrected maps over the full-width-half-maximum PB area. } 
\end{deluxetable}}


\section{Results}
\label{sec:results}
In this section we first present the results
for the individual configurations (C and D) separately, and then the results for
combined (C+D) configurations, RM corrected. We consider each step since it is helpful to
see how the configurations emphasize certain spatial scales in the source and also to see
the extent to which improvements result from the combination of these configurations.  In addition, we wish to
see the extent to which Faraday-rotation may or may not alter the magnetic field vectors for
this disturbed edge-on galaxy at C-band.  Little has been published
so far in the literature regarding wide-band polarization data and the improvements that can result from these two steps.
We finally present new CHANDRA X-ray results for NGC~2992.

\subsection{Results -- Individual Configurations}
\label{sec:results_individual}

Fig.~\ref{fig:n2992_totalintensity_DSS} shows a number of new features
in NGC~2992 at both low and high resolution.  The outer contours (D-configuration) reveal the SE extension
mentioned above, as well as extensions to the north and south; the latter appear to be offset from
the inner disk angle but do appear to align with some spiral arm structure.
The inner contours (C-configuration) also
show emission extending away from the central oval.  The `west extension'
is particularly noteworthy (see below).

The total flux density at 6.0 GHz is 80.4 $\pm$ 1.6 mJy
\citep{wie15} and the global spectral index between L-band (1.6 GHz) and C-band (6 GHz) is
$\alpha_{L-C}\,=\,-0.70\,\pm\,0.02$ \citep{li16}. This is flatter than
the majority of CHANG-ES
galaxies for which $\alpha_{L-C}\,\approx\,-0.9$, except for outliers with known AGNs
\citep[][their Fig.~4a]{li16}, i.e. NGC~2992
has a global spectral index that is suggestive of an AGN, in agreement with previous knowledge about this galaxy.
The {\it in-band}
spectral index \citep[see][for a full description]{wie15} of the highest resolution ($\sim$ 3.3 arcsec = 540 pc)
C-configuration data within the half-power-beam-width centered at the nucleus is $\alpha_{C}(nuc)\,=\,-0.78\,\pm\,0.01$.
These values clearly indicate that non-thermal emission is dominant at C-band.

\begin{figure*}
\begin{tabular}{c} 
\rotatebox{0}{\scalebox{0.77} 
  {\includegraphics[height=5truein]{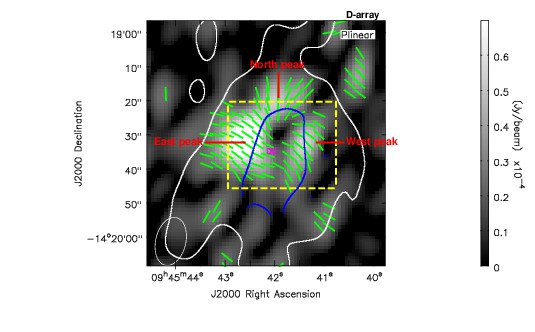}}}\\
\rotatebox{0}{\scalebox{.7}
             {\includegraphics{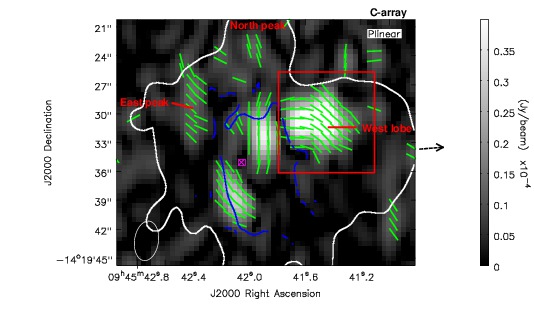}}}\\
\end{tabular}
   \caption{
     C-band linear polarization maps of NGC~2992 with apparent magnetic field vectors superimposed in green.
     {\it These vectors have not been Faraday de-rotated.} At {\bf top} 
is the D-configuration data and at {\bf bottom} is the C-configuration data (note the different map scales).  The single white contour
shows the corresponding 
total intensity 6$\sigma$ contour from Fig.~\ref{fig:n2992_totalintensity_DSS}.
The yellow dashed box (top) shows the approximate field of view that is shown in the bottom image.
The red box (bottom) shows the lobe-like feature discussed in the text.
 The interior
of the blue contours
represents where the fractional polarization falls below the believable 0.5\% level. The NED (optical) 
center of the galaxy is marked
with a magenta X in a box and the beams are shown at lower left. Several features discussed in the text
are labelled and the approximate direction of the minor axis of the galaxy is shown with a dashed arrow towards the
right in the bottom image.
}
\label{fig:n2992_polarization}
\end{figure*}

The linear polarization images 
(Fig.~\ref{fig:n2992_polarization}) are quite revealing.  The D-configuration C-band image (top) shows apparent magnetic field
(B)
vectors\footnote{These are actually electric field vectors rotated by 90 degrees.} in coherent structures, pointing
roughly away from the galaxy's center.
{\it These vectors have not yet been corrected for internal Faraday Rotation (see next section)}.
  Because the linear polarization becomes less reliable when the fractional 
polarization, $S_P/S_I$, falls below 0.5\% \citep[see][]{wie15}, we show the 0.5\% fractional polarization contour in blue. Interior to
that contour, the total intensity emission is very high so the fractional polarization is low and the results in
that central region are unreliable. Outside of the blue contour where the polarization is reliable,
systematic
behaviour is seen over size scales of $\approx$ 10 arcsec (1.6 kpc) from the nucleus.  There is emission to the
east, west and north of the nucleus which are labelled in the figure.  The west peak is at the location of
the west extension seen in the total intensity image of Fig.~\ref{fig:n2992_totalintensity_DSS}.

The lower image shows the high resolution C-configuration map.
 Again, inside the blue contour, the polarization
becomes unreliable. Thus, the two peaks closest to the galaxy's center (4 arcsec NW and SE of center)
are not believable features.  The most obvious reliable feature is
the emission in the boxed region to the west of the nucleus
 with the appearance of a radio lobe, labelled 'west lobe'. This west lobe corresponds
to the west peak labelled in the top image.  The lobe broadens somewhat at its west end.

The north and east peaks, labelled in both images, are much less obvious
in the high resolution (bottom) map. To compare, we have measured the
 flux density of the
east peak in a region
in which both maps have a percentage polarization $>0.5$\%.  The size of that region is
 1.26 $\times$
 the D-configuration beam size and 11.8 $\times$ the
C-configuration beam size. The result is 
 $S_{D-configuration}\,=\,75.3~\mu$Jy and $S_{C-configuration}\,=\,69.5~\mu$Jy, the latter
only 8\% lower. In these maps, some pixels are also blanked in the process of forming polarization
images (Sect.~\ref{sec:individual_configurations}). For C-configuration, these blanks occupy 1.5 beams (13\% of the measured
region).  If flux exists at the Q and U rms level for those pixels,
then the C-configuration flux density would increase to
$S_{C-configuration}\,=\,74.0~\mu$Jy, in close agreement with the D-configuration result.
As for the weaker north peak, it is barely detected at high resolution. The {\it maximum} brightness
 at D-configuration is 53.2 $\mu$Jy/beam which, with beam dilution (see Table~\ref{table:image_parameters}), 
would result in
4.3 $\mu$Jy/beam at C-configuration, i.e. a level of only 1.5 $\sigma$.  In summary,  
the differences between the configurations can be attributed to S/N and resolution differences
between the two observations.  The combined configuration data, presented in
Sect.~\ref{sec:combined_results}, will
make use of both of these data sets and provide
further revelations about the source structure.

\begin{figure*}
\begin{tabular}{c} 
\rotatebox{0}{\scalebox{0.6} 
             {\includegraphics[height=6truein,width=10truein]{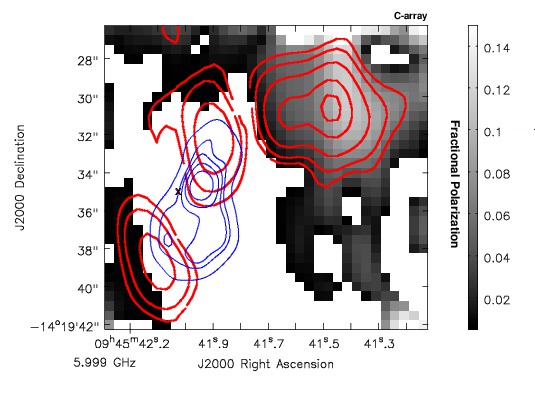}}}\\
\rotatebox{0}{\scalebox{0.6}
             {\includegraphics[height=6.5truein,width=10truein]{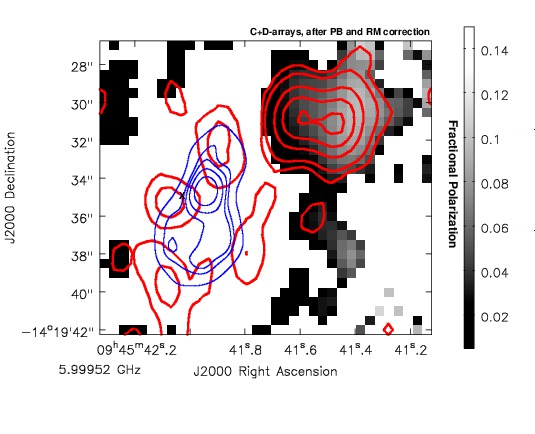}}}\\
\end{tabular}
\caption{Fractional polarization (greyscale) with linear polarization contours superimposed in
  red.  Blanked (white) regions are where the fractional polarization falls below 0.5\%.
  The C-configuration data are shown at the {\bf top} (not PB-corrected) with contours in red at
     9.6 (3$\sigma$), 15, 25, 40, and 50 $\mu$Jy/beam.
  The high resolution combined C+D configuration data (PB-corrected) are at the {\bf bottom}
with contours in red at 15, 25, 40, 50, and 58 $\mu$Jy/beam.
  Blue contours show the {\it 1.4 GHz} high resolution (1.7 x 1.2 arcsec at -17 deg) total intensity
  contours of the inner double-loop from \citet{ulv89}, at levels of
  5, 10, 12, 15, and 20 mJy beam$^{-1}$. The NED center of the galaxy is marked with a small X.
}
\label{fig:percentpol}
\end{figure*}

The west radio lobe seen at C-configuration is quite distinct.  For this feature, we have therefore
formed a map
of percentage polarization (greyscale) in
Fig.~\ref{fig:percentpol} (top)
with the unreliable regions blanked. Superimposed are red contours of linear polarization.
  The percentage polarization
  is 8.9\% at its total intensity peak (at RA = 09 45 41.48, DEC = -14 19 30.5) and shows an increase towards the west end
  of the radio lobe.
 Average values for this lobe and other parameters are listed in Table~\ref{table:results}.

 For reference, we have also downloaded from the NASA/IPAC Extragalactic Database (NED)
the high resolution {\it 1.4 GHz} total intensity image of \citet{ulv89} which shows structure associated
with the inner double-loop in the region of our central oval.  These data are represented by blue contours
in Fig.~\ref{fig:percentpol}.
This high resolution map (as well as the CHANG-ES data sets, e.g. Fig.~\ref{fig:n2992_totalintensity_DSS}) shows quite clearly
that the centre of the galaxy, as listed in NED
and taken from the optical (photographic) data of \citet{arg90} 
(RA = 09$^{\rm h}$ 45$^{\rm m}$ 42.$^{\rm \hskip -1pt s}$05,
DEC = -14$^\circ$ 19$^\prime$ 34.$^{\hskip -1pt \prime\prime}$98)
is offset towards the east of the radio core by about 1 arcsec.
The radio core is at
RA = 09$^{\rm h}$ 45$^{\rm m}$ 41.$^{\rm \hskip -1pt s}$94,
DEC = -14$^\circ$ 19$^\prime$ 34.$^{\hskip -1pt \prime\prime}$8, with an estimated error of $\pm$ 0.3 arcsec
in each coordinate.
Clearly, the radio core provides the more accurate center position, but we have retained
the NED center location mark in
our images since this was our pointing center.



\begin{figure*}
\begin{tabular}{c} 
\rotatebox{0}{\scalebox{0.65} 
             {\includegraphics[height=6truein,width=7.5truein]{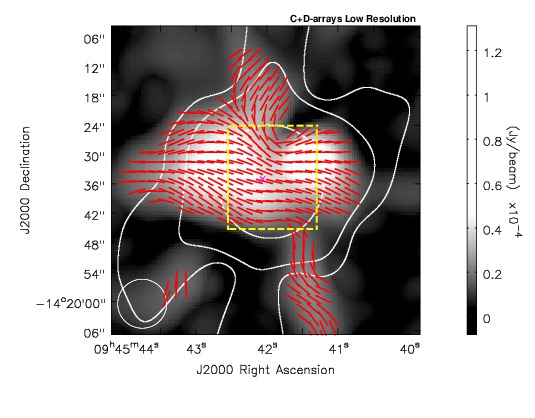}}}\\
\rotatebox{0}{\scalebox{.5}
             {\includegraphics[height=7truein,width=10truein]{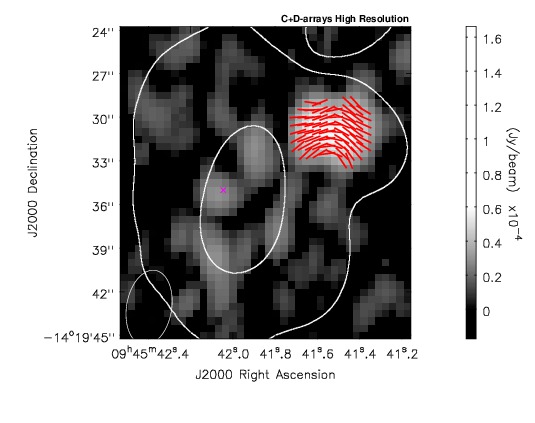}}}\\
\end{tabular}
\caption{Combined C-band C+D configuration polarization images (greyscale) with matching configuration and resolution
  total intensity contours (white) and B vectors (red, {\it Faraday corrected}). The NED center is marked
  with an X.
  {\bf Top:} Low resolution 10 x 10 arcsec image.
  Contours are 75 (5$\sigma$), 200, and 4000 $\mu$Jy beam$^{-1}$.  The peak polarized intensity is
  69.1 $\mu$Jy beam$^{-1}$ (in the west peak).
  The yellow dashed box shows the approximate field of view of the bottom image.
     {\bf Bottom:} High resolution 5 x 3 arcsec image.
     Contours are 25 (5$\sigma$), 200 and 10000 $\mu$Jy beam$^{-1}$. The maximum polarized intensity is
     60.2 $\mu$Jy beam$^{-1}$.
}
\label{fig:combined}
\end{figure*}

\begin{figure*}
\begin{tabular}{c} 
\rotatebox{0}{\scalebox{0.65} 
             {\includegraphics[height=6truein,width=7.5truein]{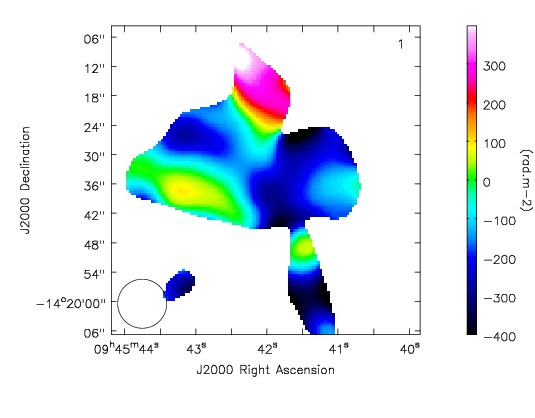}}}\\
\rotatebox{0}{\scalebox{.5}
             {\includegraphics[height=7truein,width=10truein]{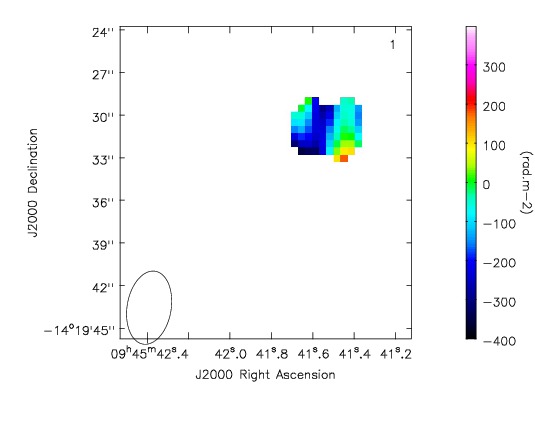}}}\\
\end{tabular}
\caption{Combined C-band C+D configuration RM images that correspond to Fig.~\ref{fig:combined} with the low resolution
  image at top and high resolution image at bottom.  The corresponding beams are shown at lower left.
}
\label{fig:combined_RM}
\end{figure*}

\begin{figure*}
\rotatebox{0}{\scalebox{0.78} 
             {\includegraphics[height=8.0truein,width=7.5truein]{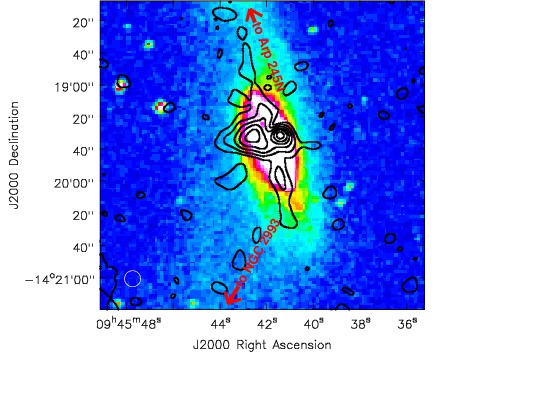}}}\\
\caption{Combined C-band C+D configuration low resolution polarization images (contours, PB-corrected) superimposed on
  an optical DSS2 blue image.  
  Contours are 10.5 (3$\sigma$), 20, 30, 40, 50, and 62 $\mu$Jy beam$^{-1}$.
  The directions towards the companion galaxies are marked and the synthesized
  beam is shown at lower left.
}
\label{fig:combined_optical}
\end{figure*}

\subsection{Results -- Combined (C+D) Configurations}
\label{sec:combined_results}

The low (10 arcsec) and high ($\approx$ 4 arcsec) resolution combined C+D configuration images are shown in
Fig.~\ref{fig:combined} (top and bottom, respectively).
The linearly polarized intensity is
shown in greyscale and several total intensity contours at matching resolution are shown in white.
The regions over which the percentage polarization
falls below believable values is similar to those shown in Fig.~\ref{fig:n2992_polarization}.
The magnetic
field orientation
{\it corrected for Faraday rotation} is now shown as red vectors. The corresponding RM maps are shown in Fig.~\ref{fig:combined_RM}.

Since the RM synthesis technique corrects for any foreground Faraday rotation, both in the galaxy
as well as Galactic, the magnetic field orientation shown in Fig.~\ref{fig:combined} is
intrinsic to NGC~2992.  The Galactic contribution to this foreground is estimated to be
-13  $\pm$ 13 
rad/m$^2$
at the location of NGC~2992
\citep{opp12}, consistent with zero.  However, taking the measurements
at face value, the Galactic contribution would be 1.8 degrees for the band center of NGC~2992 or a maximum of 2.7 degrees
for the lowest frequency.


The results discussed in Sect.~\ref{sec:results_individual} 
are still valid but now further clarity has been achieved, with the
main new results being revealed in Fig.~\ref{fig:combined} (top).
The north peak, as labelled in Fig.~\ref{fig:n2992_polarization} (top) is now seen as belonging to
emission
along the galaxy's
major axis which runs NNE to SSW (Fig.~\ref{fig:n2992_totalintensity_DSS})
and, in addition, emission is now also seen along the
SSW major axis.  Most dramatically, the west and east peaks of
Fig.~\ref{fig:n2992_polarization} are now revealed as
{\it two radio lobes on either side of the major axis} (see also Fig.~\ref{fig:combined_optical}).
As we saw before
in Fig.~\ref{fig:n2992_polarization} (bottom), the west lobe is the
more compact one and can be seen distinctly as the discrete radio lobe in Fig.~\ref{fig:combined}
(bottom).

The magnetic field vectors, once corrected for Faraday rotation,
(as they did before) show ordered fields predominantly pointing
roughly away from the galaxy's center (and AGN)
but the vectors appear to show some curvature `around' the eastern radio lobe
(Fig.~\ref{fig:combined} top).  
It is the eastern radio lobe that extends farthest from
the AGN and is not as compact or as bright as the western lobe.  It also appears to
have escaped from the galaxy's disk (see also Fig.~\ref{fig:combined_optical}).  The B vectors are
very straight in this escaping part of the eastern radio lobe.  Along the northern and southern major axes,
there is also some curvature away from the plane of the galaxy.

In the high resolution image (Fig.~\ref{fig:combined} bottom), the corrected magnetic field vectors
are seen to curve along
the western radio lobe with the curvature being somewhat more pronouced than previously
seen in Fig.~\ref{fig:n2992_polarization} (bottom).  We again produce a map of percentage polarization
for the discrete western radio lobe 
(Fig.~\ref{fig:percentpol} bottom) which, as before, shows the increase in the percentage
polarization 
towards the west end of this feature.

In summary, a comparison of Figs.~\ref{fig:n2992_polarization} and \ref{fig:combined_optical} shows that
ordered magnetic fields are appararent and the overall sense of that order is evident and consistent in both cases. However 
Faraday rotation corrections should not be neglected at C-band;
for example after correction, the magnetic field shows more curvature
in the west radio lobe.  Configuration combination also illustrates the radio lobes much more clearly.  A comparison of
the percentage polarization (Fig.~\ref{fig:percentpol})
 before (top) and
 after (bottom) configuration combination and Faraday rotation correction also shows consistent results, especially the higher percentage
 polarization (higher contribution of the ordered magnetic field) towards the end of the west radio lobe.

\subsection{X-ray Observations of NGC~2992}
\label{sec:x-ray}

Multiple CHANDRA observations were carried out for NGC~2992 with the 
Advanced CCD Imaging Spectrometer-Spectroscopy (ACIS-S) at the focus. 
One of these observations was taken without the High Energy Transmission 
Gratings (HETG). We find that the AGN in the image of this observation 
is strongly saturated. This problem is minimized in observations with 
the HETG.

\begin{figure*}
\rotatebox{0}{\scalebox{0.7} 
             {\includegraphics[height=5truein,width=10truein]{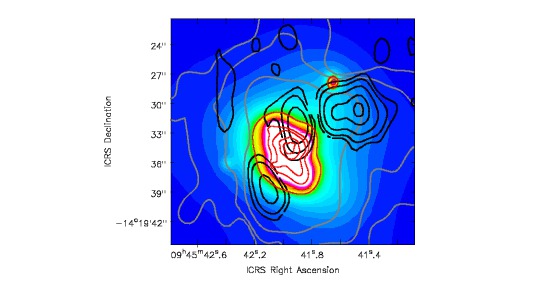}}}\\
\caption{CHANDRA X-ray image of NGC~2992 in the 0.5 - 4 keV range (colour)
 which has been adaptively smoothed with the CIAO
 routine {\sl csmooth}
 with a
Gaussian kernel whose size is
locally adjusted to achieve a  signal-to-noise ratio between 2 - 2.5.
 Grey contours are the lowest 3 contours of the C-configuration total
 intensity image as shown in Fig.~\ref{fig:n2992_totalintensity_DSS}, and black contours are the same
 contours as shown in 
 the polarized intensity image as shown in Fig.~\ref{fig:percentpol} (top).
 Four high level X-ray contours are shown in red. 
}
\label{fig:xray}
\end{figure*}

In Fig.~\ref{fig:xray} we show only an image in the 0.5-4 keV band in colour, constructed with the 
longest ACIS-S/HETG 0th-order exposure (OB \#11858) of 95 ks; the 
additional 41 ks from the other two observations would not change the 
image quality in any signficant way.
The data have been adaptively smoothed with the {\it csmooth}
routine of the CIAO (Chandra Interactive Analysis of Observations\footnote{
{\tt http://cxc.harvard.edu/ciao/}}) software analysis package.
Superimposed are C-configuration C-band total intensity contours
in grey (as in Fig.~\ref{fig:n2992_totalintensity_DSS}) and the C-configuration C-band linearly polarized contours
(as in Fig.~\ref{fig:n2992_polarization} bottom).

Closest to the galaxy's center, the X-ray emission (white part plus red contours)
is enhanced NNE/SSW which is in the direction of the galaxy's disk.
In the X-ray, the NNE/SSW orientation continues to larger scales, especially towards the SW; these features are
still along and within the confines of the optical disk as shown in Fig.~\ref{fig:n2992_totalintensity_DSS}.
The CHANDRA image clearly shows the AGN (center of the highest red contour),
and its position corresponds within errors to the radio peak
(Sect.~\ref{sec:results_individual}) rather than the NED center.

In addition to the AGN, the CHANDRA image shows the presence of the apparently diffuse X-ray emission; note that
the X-ray emission is uncorrected for the dust lane which lies approximately between the central X-ray peak and
the polarized western radio lobe.
The lack of any direct evidence for an X-ray and radio jet may indicate that the radio lobes
represent a relic of an
earlier 
episode of the  AGN (see Sect.~\ref{sec:interpretation}).


\section{Discussion} \label{sec:discussion}






The key results of this paper are that linearly polarized emission has revealed
a double-lobed radio source along the minor axis (east-west) direction of NGC~2992
that is masked in total
intensity emission, and that the western lobe is seen as a very discrete
radio lobe
at high resolution. Clearly, these polarized structures 
 have been masked by the stronger 
 unpolarized emission.

\begin{figure*}
\begin{tabular}{c} 
\rotatebox{0}{\scalebox{0.48} 
             {\includegraphics[height=6truein]{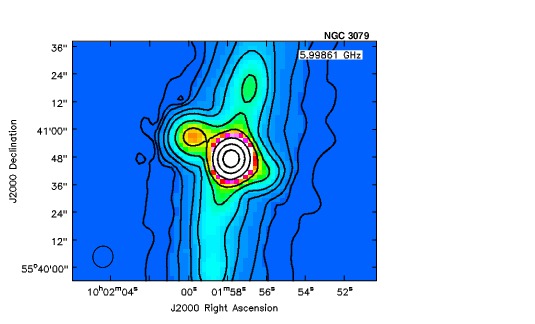}}}\\
\rotatebox{0}{\scalebox{.48}
             {\includegraphics[height=6truein]{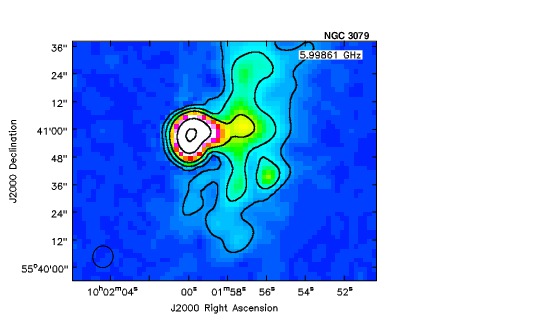}}}\\
\end{tabular}
   \caption{
     {\bf Top:} D-configuration C-band total intensity map of NGC~3079 (colours and contours).  The
contours are at 10, 25, 50, 150, 300, 700, 1300, 4000, 10000, and 20000 times the rms noise
of $6.5~\mu$Jy beam$^{-1}$.
{\bf Bottom:} D-configuration C-band linear polarization map of NGC~3079 (colours and contours). The
contours are at 10, 25, 50, 150, and 300 times the rms noise
of $5.9~\mu$Jy beam$^{-1}$.
Both maps are adapted from \citet{wie15} and the beams, 
of dimensions 9.3 by 8.7 arcsec at a position angle of  -13.9 deg., are shown at lower left. 
}
\label{fig:n3079}
\end{figure*}

There are two other good examples
in the CHANG-ES sample of known AGNs with
distinct radio 
lobes for which lobes are much more obvious in polarization than
in total intensity, namely 
NGC~3079 \citep{wie15} and NGC~4388 \citep{wie15,dam16}. 
We show
the previously unpublished total intensity and polarized
image of NGC~3079 in Fig.~\ref{fig:n3079}. Details related to
  this galaxy will be discussed elsewhere, but this is
 another example in which the outflow, in particular the radio lobe
  to the east,
 is much more obvious in polarization.  The peak total intensity 
 emission (top image) occurs at the nucleus rather than in the east lobe.
 The
brightness ratio of east radio lobe to nucleus 
(both measured at their respective peaks) is only 0.58.  
By contrast, for the linearly polarized emission (bottom), the east radio
lobe is brightest, with an east lobe to nucleus ratio of 5.8.
In other words, the contrast has increased by a factor of 10.
Similar results can be seen in NGC~4388 \citep[see][]{dam16}.
In both NGC~3079 and NGC~4388, the nuclear outflow is strong enough that the radio lobes
can also be seen in total intensity, whereas this is not the case for
NGC~2992.

\subsection{AGN-Related Outflow}
\label{sec:interpretation}

The evidence for an AGN in NGC~2992 has been outlined in Sect.~\ref{sec:n2992} and the case for AGN-related
outflow has also been made; for example \citet{vei01} indicate that the most likely energy source for the
optical outflow is a hot, bipolar, thermal wind powered by the AGN and diverted along the galaxy's minor
axis by the pressure gradient of the ISM.  In this section, we provide further evidence that the outflow
is predominantly AGN-related.

Firstly, the discrete western radio lobe seen at high resolution
in NGC~2992 (Fig.~\ref{fig:percentpol}) has the appearance
of most extragalactic radio lobes (EGRLs) seen in distant galaxies, though is small-scale in
comparison, extending to 1.6 kpc  from the galaxy's center.
It is similar to what has been seen before for Seyferts and other galaxies that
show radio lobes in total intensity (Sect.~\ref{sec:introduction}).  

Outflows are also known to result from nuclear starbursts and 
SNe in starbursts can be 
more effective at driving galactic-scale winds than AGN \citep{str04} in LLAGNs.
Such winds
are dominated by hot gas, visible in X-ray emission
and in H$\alpha$ \citep[e.g.][and references therein]{hee11}.
However, these winds (such as are known)
are in conical ouflows.  They also
align with the galaxies' minor axes,
{\it even at the smallest scales}, i.e. as soon as the outflow leaves the SF region,
 because the alignment is a result of the surrounding
ISM which focusses the outflow and causes it to move in the direction of steepest pressure gradient
\citep[e.g.][]{str04}.
For the starburst-driven wind in NGC~253, both the total intensity as well as the polarized
radio emission are also
filamentary and surround the nuclear outflow which is collimated by a helical magnetic field
  into a wide-angle cone \citep{hee11}.
 In M~82, the radio emission is also observed
along the outside of the wind \citep{ade13}.  In general, starburst-related winds have
opening angles of
10 to 45 deg. near the base increasing to 45 to 100 deg. above the disk
\citep{vei05} for those systems that are well-known.


The distinct western lobe in NGC~2992, by contrast, appears more `lobe-like' than
 `cone-like' in these observations.
It is also {\it mis-aligned} with the galaxy's center.  Fig.~\ref{fig:percentpol} illustrates
this best in which we have shown
high resolution 1.4 GHz contours from the inner double-loop 
\citep{ulv89} which is in our inner central oval location.
The orientation of the inner double-loop is strongly misaligned
with the minor axis and is well-known to represent ejection from the AGN.
The western radio lobe then becomes aligned
with the minor axis.
 These observations suggest that outflow near the nucleus 
first extends towards
the NNW and then it curves towards the west (the direction of the minor axis)
as expected from ISM pressure gradients.
On larger scales towards the east and west
(Fig.~\ref{fig:combined_optical} the alignment is roughly along the minor axis.

Ordered magnetic fields are also typical of aligned outflows from AGNs.
In EGRLs,  magnetic field alignments and their causes are still debated
\cite[e.g.][]{tuc12}. However, in classic aligned outflow, 
the radial component of the field would fall off as $1/r^2$, since flux is conserved
and the radial component declines with increasing cross-sectional area.
A helical field wrapping around the outflow would
decline as $1/r$, that is, as the circumference.
Consequently, one would expect B to align parallel to the jet close to the origin, as
many such observations show \citep{caw93,gom08}, but
once the circumferential field starts to dominate, B would become perpendicular
farther from the origin, as is sometimes, but not always
observed \citep{sch00}.  
Alternatively, $B$ 
parallel to the outflow axis may result if the initial fields are sheared to lie in a plane.
Strong shocks can compress an initially random field,
creating an ordered magnetic field, in which case transverse magnetic field vectors would
result in the region of the shocks \citep{tuc12}. 
Radio knots are generally interpreted as the termination point of a jet that is
stopped in the dense ISM where the radio emission is enhanced by compressed material
and magnetic fields \citep{fal98,lei06}. 

In the case of the west lobe of NGC~2992 
(Fig.~\ref{fig:combined} bottom), we observe curving magnetic fields.  The curvature
appears to follow the direction that the radio lobe, itself, would be bending because of pressure gradients in
the ISM.  The fields become somewhat more transverse at the west (terminating) end, similar
to the EGRLs just described. An 
increase in percentage polarization towards the end of the lobe
(Fig.~\ref{fig:percentpol}) is consistent with this 
interpretation\footnote{Note that
\citet{han09} show that organized 
magnetic fields can be set up in galaxies within 2 Gyr of disk evolution
from initially randomly oriented 
exploding magnetized stars; that is, organized fields can occur without
any outflow. However, such fields are on
a global scale and cannot explain embedded structures such as are observed here.}.

From optical emission lines, 
 there is also much evidence for radial flows including cone-like 
outflow \citep{vei01} towards the east and west. Our total intensity image
(Fig.~\ref{fig:n2992_totalintensity_DSS}) as well as our 
low-resolution polarization images that show magnetic fields pointing radially away from the nucleus
(Fig.~\ref{fig:combined} top) agree with this picture that on larger scales, we can see wider 
outflow.  A detailed comparison of the optical and radio outflows is beyond the
scope of this paper.  However, we do note that
the distinct west radio lobe is located within the western optical cone in projection. 
Finally, \citet{fri10} have shown, on energetic grounds, that outflows on scales up to 2 kpc 
cannot have been produced by a starburst and must have originated with the AGN. This is approximately
the same spatial scale over which we see the radio lobe.

In summary, the weight of evidence indicates that we are observing embedded weak polarized radio lobes
that originate with an AGN in NGC~2992.

\subsection{Interaction with the ISM and Physical Parameters}
\label{sec:interaction}

It is clear that the radio lobes are interacting with the ISM.  On the west side,
the lobe has not yet escaped from the galaxy but the east lobe has emerged
outside of the disk.  The discrete west
 lobe of NGC~2992, as discussed above, does not align directly with the nucleus
 (Fig.~\ref{fig:percentpol}) and, if the AGN-related outflow has maintained its direction
 over time, then there has been a significant bend in the outflow direction 
 between 3 and 7 arcsec from the nucleus ($\approx$ 500 - 1200 pc in projection). 
Such bends are also commonly seen in EGRLs, for a variety of
reasons (e.g. precessing jets, Kelvin-Helmholtz 
instabilities, etc.) but for NGC~2992, the bend is more likely caused by
a deflection due to pressure gradients in the ISM as described 
in early models such as \citet{hen81, fie84}.  Similar bending has
been observed before in spiral galaxies, e.g. in NGC~4258 \citep{pla91,kra07}.

Measured and derived properties of the distinct western radio lobe
of NGC~2992
are listed in Table~\ref{table:results} 
measured within the 3$\sigma$ contour 
as shown in Fig.~\ref{fig:percentpol} (top).

The percentage polarization is significant, i.e. $\approx$ 6\%.
This value may, however, be underestimated.  This is because we do not
observe the radio lobes in total intensity and have simply measured the total intensity over a region that
corresponds to the observed polarized radio lobe.  Consequently, the total intensity measurement could include any
other total intensity emission along the line of sight that is outside of the radio lobe itself, thus increasing
$S_I$ in comparison to $S_P$\footnote{Free-free emission is also present but
  is a minor contributor to total intensity in comparison with non-thermal emission.}.
As a comparison,
in distant EGRLs, values are only a few percent
 at this frequency 
\citep[][and references therein]{tuc12}, possibly because of poorer linear resolution for the more
distant sources (but see also \citet{bag07}).
However, higher percentages have been seen in other nearby spiral galaxies,
for example the similar galaxy, NGC~4388
\citep{dam16}. 

In Table~\ref{table:results}, we also provide the minimum energy magnetic
field\footnote{Although we cannot verify that the minimum energy criterion strictly applies, it is the
  only way to obtain such quantities and presents a baseline for possible future comparisons.};
 this was determined
from the total
intensity flux density, following \citet{bec05}, assuming that the measured
spectral index $\approx$ the non-thermal spectral index.  If the spectral index steepens
by 10\% due to a thermal contribution,
then $B_{min}$ increases by
+4\%.
If the line of sight distance decreases by a factor of 2, then  $B_{min}$ will increase by
18$\to$20\%.  
The same cautions outlined in the previous paragraph apply here.

The minimum total field value (Table~\ref{table:results}), if random,
suggests a magnetic pressure of 
$5\,\times\,10^{-11}$ erg cm$^{-3}$. By comparison, typical values for
 galactic disks are
 $~10^{-11}$ erg cm$^{-3}$ \citep[e.g.][]{beck07,bec15}
and are about 3 times larger in the central regions.
Consequently, the value is high enough that the jet should propagate but
not so high that it wouldn't be affected by the surrounding ISM, as the changes in
orientation suggest.  As further confirmation,  \citet{col05} indicate that, within the central
5 arcsec, the pressure of the X-ray emitting gas is ${\phi_X}^{-1/2} (0.5 - 3)\times 10^{-11}$ erg cm$^{-3}$,
where ${\phi_X}$ is the volume filling factor, and they also indicate that this value
is similar to the extended narrow line region clouds.  
Consequently, pressure calculations from a wide variety of observations are all pointing
to values that are within factors of a few, of order several 10$^{-11}$ erg cm$^{-3}$.

With sufficient data, it should be possible to make use of such observations to determine
(or place limits on) the
physical conditions in a galaxy, or the energetics of the outflow, or both.  For example,
the `stall' location of a radio lobe should occur where the lobe pressure equals the ISM pressure,
and a lobe should emerge from the disk if its pressure exceeds that of the ISM (as for the
east lobe of NGC~2992).  The relation between the intrinsic bending angle
and jet ram pressure is also known for environments with static atmospheres \citep[][]{hen81,irw88}.
Since NGC~2992 has been well-observed at a variety of wavelengths (Sect.~\ref{sec:n2992}),
this galaxy should be well-suited
to the development of a detailed model that includes the X-ray, optical and other data that are
available.


{
\begin{deluxetable}{lcccccccc}
\hspace*{-10cm}
\tabletypesize{\tiny}
\renewcommand{\arraystretch}{1.1}
\tablecaption{Parameters of the NGC~2992 West Radio Lobe\tablenotemark{a}\label{table:results}}
\tablewidth{0pt}
\tablehead{
\colhead{Observation}  
& \colhead{$S_I$\tablenotemark{b}} 
& \colhead{$S_P$\tablenotemark{b}}
& \colhead{$SP_I$\tablenotemark{c}} 
& \colhead{$SP_P$\tablenotemark{c}}
& \colhead{$\overline{P/I}$\tablenotemark{d}} 
& \colhead{$\overline{\alpha}$\tablenotemark{e}} 
& \colhead{${B_{min}}$\tablenotemark{f}}
& \colhead{${\rm P}_B$\tablenotemark{g}}
\\
 & (mJy) & ($\mu$Jy) & ($\times$10$^{19}$ W Hz$^{-1}$) & ($\times$10$^{19}$ W Hz$^{-1}$)  &  (\%) &
 & ($\mu$G) & ($\times$10$^{-11}$ erg cm$^{-3}$) \\ 
}
\startdata
{\bf West lobe}
&  2.21 $\pm$ 0.03 & 97 $\pm$ 2 & 30.6 $\pm$ 0.4
& 1.34 $\pm$ 0.03
&
6.4 $\pm$ 0.2 & -0.85 $\pm$ 0.16 & 35.6 & 5.0\\
\enddata
\tablenotetext{a}{Measurements for the radio lobes seen in
  the C-configuration image of Fig.~\ref{fig:percentpol} (bottom).  All measurements apply to the
regions within the 3$\sigma$ contours, and where the fractional polarization has an unblanked value.}
\tablenotetext{b}{Flux densities of the total intensity emission
and the linearly polarized emission.  The uncertainty estimate reflects variations that result from
varying the irregularly sized measurement region, whose value dominates over the error calculated from
the rms noise values given in Table~\ref{table:image_parameters}.}
\tablenotetext{c}{Spectral power of the total intensity and polarized emission.}
\tablenotetext{d}{Average fractional polarization measured from Fig.~\ref{fig:percentpol}.}
\tablenotetext{e}{Average total intensity spectral index ($S_\nu\propto \nu^{\alpha}$) and its error from maps 
(not shown) generated
as described in \citet{wie15}.}
\tablenotetext{f} {Minimum energy (total) magnetic field computed according to the prescription of
\citet{bec05} assuming the number density ratio of protons to electrons
per unit energy, $K_0=100$, the line of sight
distance is equal to the width of the feature, and using values of $S_I$ and $\alpha$ in this table.
See Sect.~\ref{sec:discussion} for estimates of some uncertainties.}
\tablenotetext{g} {Magnetic pressure, assuming random fields (${\rm P}_B\,=\,{B_{min}}^2/8\pi$).}
\end{deluxetable}}

\subsection{A Possible Evolutionary Scenario}

The simplest scenario that is consistent with the data presented here
(although it is not a unique interpretation), is that the
AGN outflow in NGC~2992 has been episodic but not changing direction at the source, or that episodic outflows
are embedded within steadier low level emission.

For example,
the most recent activity would belong to the total intensity inner double-loop region within our inner oval
region (Fig.~\ref{fig:n2992_totalintensity_DSS}) and oriented NNW-SSE.
We would not see this outflow explicitly in linear polarization
because the percentage polarization is below believable levels near the galaxy's center.

At an earlier time, the higher resolution western radio lobe observed in linear polarization was formed
and the observed polarized feature would therefore be a relic of this earlier activity. 
Pressure gradients in the ISM cause this outflow to bend towards the minor axis.
The low resolution eastern and western radios lobe could have occurred at approximately the same time or else
they represent lower level continuous flows (e.g. note that the B vectors don't curve around the lobes as is sometimes
seen in EGRLs).
Since NGC~2992 is
highly disturbed because of the interaction with its companions, the eastern disk could have a
different ISM pressure gradient, allowing the eastern lobe to escape from the disk as observed.
These features would not be seen in total intensity because the broader scale total intensity emission is
much brighter, masking the polarized intensity structure.

Finally, the southeast extension
(Fig.~\ref{fig:n2992_totalintensity_DSS}) would belong to an even earlier episode.  This feature has emerged
from the galaxy and is pointing towards the companion galaxy, NGC~2993. Polarized emission would not be seen
from this extended feature because it is just too weak on this larger scale.

Since the optical emission appears as a wide-angled cone \citep{all99} with the west radio lobe within it in
projection, the discrete radio features may represent specific high-energy ejection eposides in comparison to
lower level steady emission.  The latter is more likely explained by an AGN-powered thermal wind as described in
\cite{vei01}.

\section{Conclusions}
\label{sec:conclusions}

The polarized emission from NGC~2992,
uniquely provided by the CHANG-ES survey, has offered us 
a new glimpse into the nuclear activity in this galaxy. Although radio lobes are {\it not}
observed in total intensity radio emission, they {\it are} observed in polarized emission.  Thus,
we are observing weak radio lobes which were not strong enough to be distinguished from
other non-polarized emission seen in total intensity.

For NGC~2992, at high resolution, a single discrete radio lobe is observed on the western side of
the galaxy's center and at low resolution, two radio lobes are seen on either side of the nucleus,
extending to the east and west.  The eastern radio lobe has emerged from the galaxy whereas the
western lobe is within the disk.
We argue that the radio lobes are consistent with AGN-like outflow.  The orientation of the discrete
western radio lobe is consistent with outflow that has been redirected, via pressure-gradients in the
ISM, along the minor axis.  The outflow and ISM pressures are consistent with this
picture, to within measurable factors of a few (assuming that the minimum energy criterion holds).

We suggest a possible (though not unique) evolutionary scenario for the observed radio features such that
the variety of radio activity seen in this galaxy can be simply explained by episodic
outflow from the AGN, possibly embedded within more steady lower level flows.
The AGN is at the radio core position which
should designate the center of the galaxy, rather than the normally quoted optical center. 

Such polarized radio lobes provide
a unique opportunity to study the physics of outflow/ISM interactions in spiral galaxies with LLAGNs.



\acknowledgments

This work has been supported by a Discovery Grant to the first author by the Natural Sciences and Engineering Research
Council of Canada. 
This research has made use of the NASA/IPAC Extragalactic Database (NED) which is operated by the Jet Propulsion Laboratory, California Institute of Technology, under contract with the National Aeronautics and Space Administration. 
The National Radio Astronomy Observatory is a facility of the National Science Foundation operated 
under cooperative agreement by Associated Universities, Inc. We thank Dominic Schnitzeler for his helpful comments to improve this paper.

{\it Facilities:} \facility{VLA}.

\end{document}